\documentclass[11pt,draftcls,onecolumn]{IEEEtran}
\usepackage{graphicx}
\usepackage{caption}
\usepackage{subfigure}
\usepackage{amsmath}
\usepackage{amssymb}
\usepackage{bm}
\usepackage{color}
\usepackage{url}

\graphicspath{{./fig/}}

\begin{document}
\title{{\huge Application of Smart Antenna Technologies in Simultaneous Wireless Information and Power Transfer} }
 \markboth{\textit{A Manuscript Submitted to The IEEE     Communications Magazine} }{}
\author{Zhiguo Ding, Caijun Zhong,  Derrick Wing Kwan Ng, Mugen Peng,   Himal A. Suraweera, Robert Schober and H. Vincent Poor
\thanks{
Z. Ding, M. Peng, and   H. V. Poor are with the Department of
Electrical Engineering, Princeton University, Princeton, NJ 08544,
USA. Z. Ding is also with the School of Computing and Communications, Lancaster University, LA1 4WA, UK.   M.
Peng is also with the Key Laboratory of Universal Wireless
Communications for the Ministry of Education, Beijing University of
Posts and Telecommunications, Beijing, China.

D. W. K. Ng and R. Schober are with the Institute for Digital
Communications, University of Erlangen-Nurnberg, Germany.

C. Zhong is with the Institute of Information and Communication
Engineering, Zhejiang University, China.

H. A. Suraweera is with the Department of Electrical and
Electronic Engineering, University of Peradeniya, Peradeniya
20400, Sri Lanka.

  }
}

\date{\today}
 \maketitle

\begin{abstract}
Simultaneous wireless information and power transfer (SWIPT) is a
promising solution to increase the lifetime of wireless nodes and
hence  alleviate the energy bottleneck of energy constrained
wireless  networks. As an alternative to conventional   energy
harvesting techniques, SWIPT relies on the use of   radio
frequency signals, and  is expected  to bring some fundamental
changes to the design of wireless communication networks. This
article  focuses  on the application of   advanced smart antenna
technologies, including multiple-input multiple-output   and
relaying
 techniques, to SWIPT. These
smart antenna technologies have the potential to significantly
improve the energy efficiency and also the spectral efficiency of
SWIPT. Different network topologies with single and multiple users
are investigated, along with some promising solutions to achieve a
favorable   trade-off between system performance and complexity. A
detailed discussion of    future   research challenges for the
design of SWIPT systems is also provided.
\end{abstract}
\newpage

\section{Introduction}
In wireless power transfer, a concept originally conceived  by
Nikola Tesla in the 1890s, energy is transmitted  from a power
source to a destination  over   the wireless medium.
The use of wireless power transfer can avoid the costly process of
planning and installing power cables in buildings and
infrastructure.  One of the challenges for implementing  wireless power
transfer is its low energy transfer efficiency, as only a small fraction of
the emitted energy can be harvested at the receiver due to
severe   path loss  and the low efficiency of radio frequency (RF) - direct current (DC) conversion. In addition,   early   electronic devices,
such as first generation mobile phones, were bulky and suffered
from high power consumption. For the aforementioned  reasons,
wireless power transfer had not received much attention until
recently, although Tesla had already  provided a successful
demonstration to light  electric lamps wirelessly in 1891.

In recent years, a significant amount of  research effort has
been dedicated   to reviving  the old ambition of wireless power
transfer, which is motivated by the following two reasons
\cite{LV:SWIPT08,JR:MIMO_WIPT_Rui_Zhang}. The   first reason
is the tremendous  success of wireless sensor networks (WSNs)
which have been widely applied for  intelligent transportation,
environmental monitoring, etc. However, WSNs are energy
constrained, as each sensor has to be equipped with a battery
which has a limited lifetime in most practical cases. It is often
costly to replace these batteries and the application of
conventional energy harvesting (EH) technologies relying on natural
energy sources is problematic due to their intermittent nature.
Wireless power transfer can be used as a promising alternative to
increase the lifetime of WSNs.     The  second reason is  the now
widespread use of low-power devices that can be   charged
wirelessly. For example, Intel has demonstrated the wireless
charging of a temperature and humidity meter as well as a
liquid-crystal display using the signals of a  TV
station $4$ km away \cite{JR:INTEL}.

This article   considers the combination of wireless power
transfer and information transmission, a recently developed
technique termed {\it simultaneous wireless information and power
transfer} (SWIPT), in which information carrying signals are also
used for energy extraction. Efficient SWIPT requires some
fundamental changes in the design of wireless communication
networks. For example, the conventional criteria for evaluating
the performance of   a wireless system are the information
transfer rates and the reception reliability. However,
if  some users in the system   perform  EH by using
RF signals, the trade-off between the achievable
information rates and the amount of   harvested energy becomes an
important figure of merit \cite{LV:SWIPT08}. In this context, an
ideal receiver, which has the capability to perform information
decoding (ID) and EH simultaneously, was considered in
\cite{LV:SWIPT08}. In \cite{JR:MIMO_WIPT_Rui_Zhang}, a more
practical receiver architecture was proposed, in which the receiver
has two circuits to perform ID and EH separately.

This article focuses on the application of smart antenna
technologies, namely multiple-input multiple-output (MIMO) and
relaying, in SWIPT systems. The use of these smart antenna
technologies is motivated by the fact that they have the potential
to improve the energy efficiency of wireless power transfer
significantly.  For example, MIMO can be used to increase the lifetime of  energy constrained sensor networks, in which a data fusion center is equipped with multiple antennas with which it can focus  its RF energy on sensors that need to be charged wirelessly,   leading to a more energy efficient
solution compared to a  single-antenna transmitter. Furthermore, a
relay can harvest energy from RF signals from a source   and then
use the harvested energy to forward information to the
destination,  which not only facilitates the efficient use of RF
signals but  also provides motivation for information and energy cooperation among
wireless nodes \cite{zhengcr}. The application of smart antenna technologies to
SWIPT opens up many new exciting possibilities but also brings
some challenges for improving spectral and energy efficiency in
wireless systems.   The organization of this  article is as
follows. Some basic  concepts of SWIPT are introduced first. Then,
the separate and joint application of MIMO and   relaying in SWIPT
is discussed in detail. Finally    some future research challenges
for the design of multi-antenna and multi-node SWIPT systems are
provided.

\section{SWIPT: Basic Receiver Structures}
\begin{figure*}[t]
\centering
\includegraphics[width=5in]{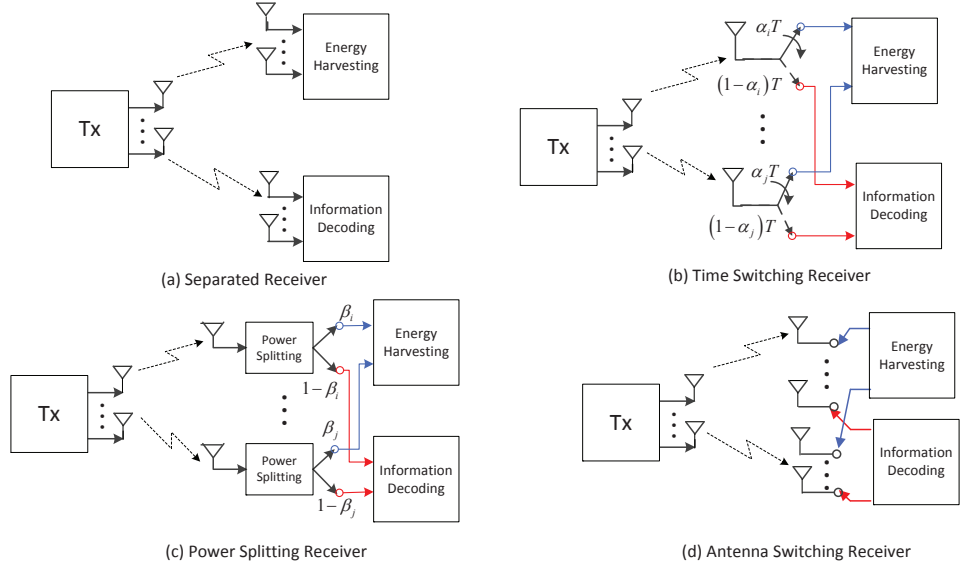}
\caption{\footnotesize Illustration of  the described SWIPT
receiver structures. $\alpha_i$ denotes the time switching factor,
$\beta_i$ denotes the power splitting factor, $i$ denotes the antenna index, and $T$ denotes the
  transmission block duration.  } \label{fig:cap_sys}
\end{figure*}

In SWIPT systems, ID and EH cannot be performed on the same received signal in
 general.   Furthermore,
   a receiver with a single antenna typically may not be able to
collect enough energy to ensure reliable power supply. Hence,
centralized/distributed antenna array deployments, such as MIMO
and relaying, are required to generate sufficient power for
reliable device operation.   In the following, we provide an
overview
  of MIMO SWIPT  receiver structures, namely  the power splitting, separated,
   time-switching, and antenna-switching  receivers, as shown in Fig. \ref{fig:cap_sys}.
\subsection{Separated Receiver} In a separated receiver architecture, an EH circuit and an ID circuit
are implemented into two separate receivers with separated
antennas,   which   are
served by a common multiple antenna transmitter
\cite{JR:MIMO_WIPT_Rui_Zhang}. The separated receiver structure
can be easily implemented using off-the-shelf components for the
two individual receivers. Moreover, the trade-off between the achievable information rate and the harvested energy
can be optimized based on   the channel state information (CSI) and feedback  from the two individual receivers to the transmitter. For instance, the covariance matrix of the transmit signal
 can be optimized for capacity maximization of the ID receiver subject to a minimum required amount of energy transferred to the EH receiver.

 \subsection{Time Switching Receiver}  This receiver consists of an information decoder, an RF energy harvester, and a switch at each antenna \cite{JR:MIMO_WIPT_Rui_Zhang}.
  In particular, each receive antenna can switch between the EH circuit and the ID circuit periodically based on a time switching sequence for EH
  and ID, respectively. By taking into account the channel statistics and the quality of service requirements regarding the energy transfer, the time
  switching sequence and the transmit signal can be jointly optimized for different system design objectives.

 \subsection{Power Splitting Receiver} Employing a passive power splitting unit, this receiver splits the received power at each antenna into
two power streams with a certain power splitting ratio before any
active analog/digital signal processing is performed. Then, the
two streams are sent to an energy harvester and an information
decoder, respectively, to facilitate simultaneous EH and ID
\cite{JR:MIMO_WIPT_Rui_Zhang,JR:Kwan_secure_imperfect,JR:EE_SWIPT}.
The power splitting ratio can be optimized for each receive
antenna. In particular, a balance can be struck between the system achievable information rate   and the harvested energy by varying the value of the power
splitting ratios. Further performance improvement can be
achieved by jointly optimizing the signal and the power splitting
ratios.

\begin{figure*}[t]
\centering
\includegraphics[width=5in]{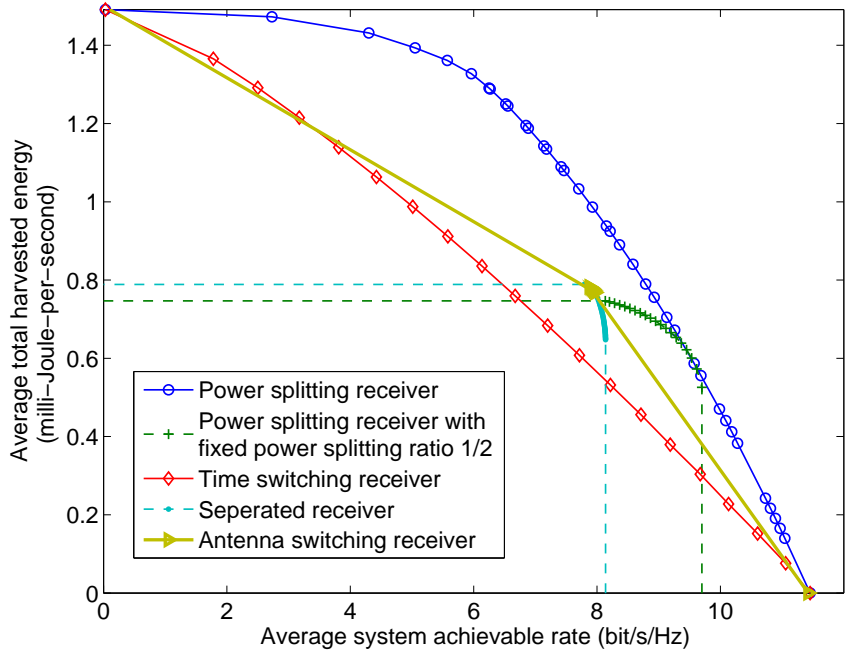}
\caption{\footnotesize The trade-off region of the average total
harvested energy (mJ/s) and the average system achievable rate
(bit/s/Hz)   for the different receivers. The carrier frequency is
$915$ MHz and the  receiver is located $10$ meters from the
transmitter. The total transmit power, noise power, transceiver
antenna gain, and RF-to-electrical energy conversion loss are set
to $10$ Watt, $-23$ dBm, $10$ dBi, and $3$ dB, respectively. The  multipath fading coefficients are modelled as
independent and identically distributed Rician random variables
with a Rician $K$-factor of $6$ dB. } \label{fig:cap_EH}
\end{figure*}

\subsection{Antenna Switching Receiver }
 With multiple
antennas, low-complexity antenna switching between
decoding/rectifying can be used to enable SWIPT \cite{I.Krikidis}.
For instance, given $N_R$ antennas, a subset of $L$ antennas can
be selected for ID, while the remaining $(N_R-L)$ antennas are
used for EH. Unlike the time switching  protocol which requires
stringent time synchronization and the power splitting  protocol
where performance may degrade in case of  hardware imperfections,
the antenna switching protocol is easy to implement, and
attractive for practical
 SWIPT designs. From a theoretical point of view,   antenna switching may be interpreted as
 a special case of power splitting with binary power splitting
ratios at each receive antenna.

Fig. \ref{fig:cap_EH} illustrates the performance trade-offs of
the considered SWIPT receiver structures \cite{JR:MIMO_WIPT_Rui_Zhang}. In particular, we show
the average  total harvested energy versus the average system
achievable information  rate  in a point-to-point scenario with one transmitter and one receiver.   A transmitter equipped with
$N_{\mathrm{T}}=2$  antennas  is serving a receiver equipped  with
$N_{\mathrm{R}}=2$ receive antennas. Resource allocation is
performed to achieve the respective optimal system performance in
each case \cite{Xlu14}. For a fair comparison, for the separated receiver, the
EH receiver and the ID receiver are equipped with a single
antenna, respectively, which results in $N_{\mathrm{R}}=2$.
Besides, we also illustrate  the trade-off region for a suboptimal
power splitting receiver with a fixed power splitting ratio of
$\frac{1}{2}$ at each antenna.   It can be observed that the optimized power
splitting receiver achieves the largest trade-off region among
 the considered receivers at the expense of incurring the highest hardware
complexity and the highest computational burden for resource
allocation.

\section{ MIMO SWIPT Networks} \label{section: MIMO SWIPT}

\begin{figure*}[t]
\centering
\includegraphics[width=5in]{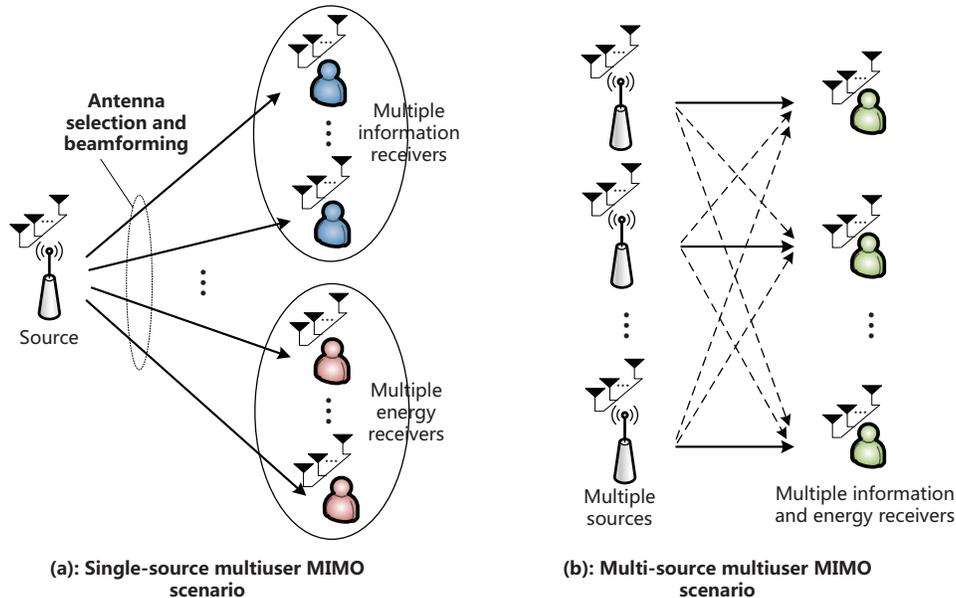}
\caption{\footnotesize Two typical multiuser MIMO scenarios.}
\label{fig1}
\end{figure*}

MIMO can be exploited to bring two distinct benefits to SWIPT networks. On  the one
hand, due to the broadcast nature of wireless transmission, the
use of  additional antennas at the receiver  can yield more
harvested energy. On the other hand, the extra transmit antennas
can be exploited for beamforming, which could significantly
improve the efficiency of information and energy transfer. The
impact of MIMO on point-to-point SWIPT scenarios with one source,
one EH receiver, and one ID receiver was studied
in \cite{JR:MIMO_WIPT_Rui_Zhang}, where  the   trade-off between the MIMO information
rate and power transfer was  characterized.  The benefits of
MIMO are even more obvious  for the multiuser MIMO scenario
illustrated in Fig. \ref{fig1}(a). Specifically,  a source
equipped with multiple antennas serves multiple information
receivers, where  the RF signals intended for the ID
receivers can also be used to charge  EH receivers wirelessly.
Since there are   multiple users in the system, co-channel
interference (CCI)  needs to be taken into account, and various
interference mitigation strategies can be incorporated into SWIPT
implementations, e.g. block diagonalization precoding as in
\cite{BD}, where information is sent to receivers that are
interference free, and energy is transmitted to the remaining
receivers. Furthermore, it is beneficial to employ user scheduling,
which allows receivers to switch their roles between an EH
receiver and an ID receiver based on the channel quality
in order to further enlarge the trade-off region between the information
rate and the harvested energy.


The multi-source multiuser MIMO scenario illustrated in Fig.
\ref{fig1}(b) is another important SWIPT application, where
multiple source-destination pairs share the same spectrum and the associated
interference control is challenging. Since in interference channels, interference signals and
information bearing signals co-exist,
issues such as interference collaboration and coordination bring both new
challenges and new opportunities for the realization of SWIPT, which are
very different from those in the single source-destination pair scenario. For example,
with   antenna selection and interference alignment as illustrated
in \cite{IA}, the received signal space can be partitioned into two
subspaces, where the subspace containing the desired signals is  used for
information transfer, and the other subspace containing the  aligned interference is
used for power transfer. This design  is a win-win strategy since
the information transfer is protected from interference, and the
formerly discarded interference can be utilized as an energy source.
More importantly, this approach  offers a new look at   interference
control, since the formerly undesired and useless interference can be used to enhance the performance of   SWIPT systems. On the other hand,   the use of RF EH introduces additional  constraints to the design of transmit beamforming. Hence,  the  solutions well-known from  conventional wireless networks, such as zero forcing and maximum ratio transmission, need to be suitably modified to be applicable in  SWIPT systems, as shown in \cite{IM}.

\section{Relay Assisted SWIPT  Systems}\label{section: Relay MIMO}
\begin{figure*}[!h]
\centering
\includegraphics[width=5in]{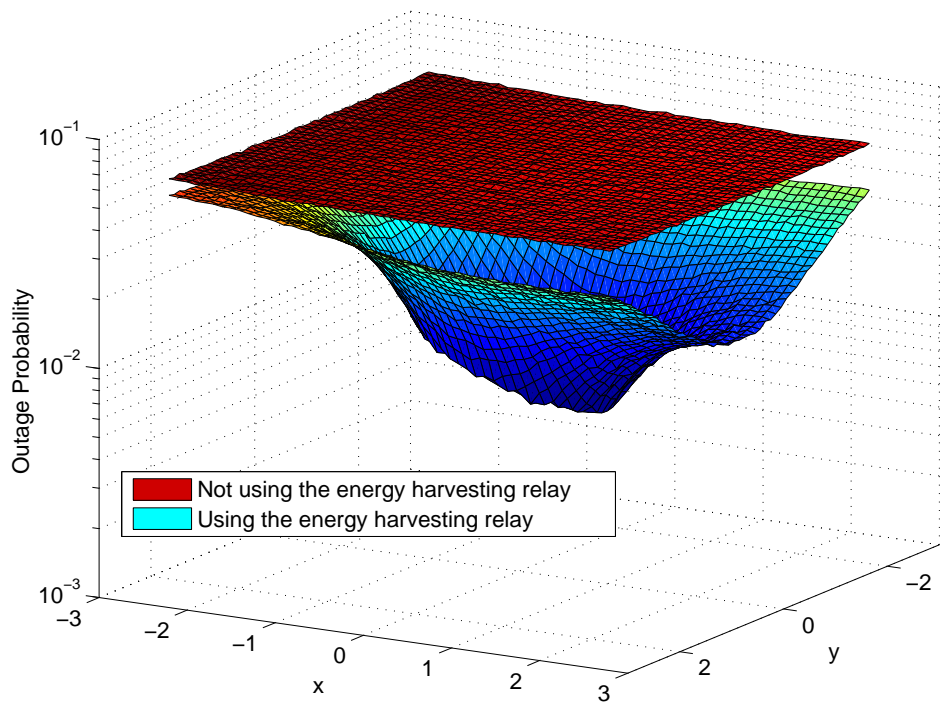}
\caption{\footnotesize Outage performance of a relaying network
with one source, one relay, and one destination. The source is
located at $(0,0)$, the destination is located at ($5$ m,0), and
the x-y plane shows the location of the relay.  The carrier
frequency is $915$ MHz. The total
transmit power, noise power, transceiver antenna gain, and
RF-to-electrical energy conversion loss are set to $10$ Watt, $-17$
dBm, $0$ dBi, and $3$ dB, respectively. We assume that the
multipath fading coefficients are modelled  as independent and
identically distributed Rayleigh random variables. The targeted
data rate is $0.1$ bit/Hz/s. The path loss exponent  is $3$.  }
\label{fig:location}
\end{figure*}

Centralized MIMO as described in Section \ref{section: MIMO SWIPT}
may be difficult to   implement due to   practical constraints,
such as the  size and cost of  mobile devices. This motivates the
use of   relaying   in SWIPT networks. In addition,
the use of wireless power transfer will encourage  mobile nodes to
participate in cooperation, since relay transmissions can be
powered by the energy harvested by the relay from the received RF signals and
hence the battery lifetime  of the relays can be increased.   The
benefits of using   EH  relays can be illustrated
based on  the following example. Consider a relaying  network with
one source-destination pair and a single decode-and-forward (DF)
relay.   SWIPT is performed at the relay by using the power
splitting receiver structure   shown in Fig. \ref{fig:cap_sys}.   The performance of the scheme
using   this    EH relay is compared to that of direct
transmission, i.e., when the relay is not used,  in Fig.
\ref{fig:location}.  As can be observed from the figure, the use
of an  EH relay can decrease the outage probability
from $7\times 10^{-1}$  to $5\times 10^{-2}$, a more than ten-fold
improvement in reception reliability, compared to direct
transmission.

 The performance of time sharing and power splitting SWIPT systems employing     amplify-and-forward (AF) and DF relays was analyzed in \cite{A.Nasir}, and the impact of power allocation   was investigated in
 \cite{relay_selectionding2}. These existing results demonstrate
 that    the behavior of the outage probability in relay assisted SWIPT systems is different from that in conventional   systems  with self-powered relays.   For example, in the absence of a direct source-destination link, the outage probability with an EH relay decays with increasing   signal-to-noise ratio (SNR) at a
 rate of $\frac{\log SNR}{SNR}$, i.e.,  slower than the rate of $\frac{1}{SNR}$    in conventional systems. The reason for this   performance
  loss is that   the relay transmission power
   fluctuates with the source-relay channel conditions.  This  performance loss can be mitigated
    by exploiting  user cooperation. For example,  in a network with multiple user pairs and an EH relay, advanced power allocation strategies, such as  water filling based and auction based
     approaches,
        can be used  to ensure that  the outage probability  decays  at the faster rate of $\frac{1}{SNR}$  \cite{relay_selectionding2}. This performance gain is obtained because   allowing user pairs to share power can avoid the situation in which  some users are lacking   transmission
         power whereas the others have more power than needed.

Relay selection is an important means to exploit multiple relays with low system complexity, and the use of EH   also brings   fundamental
changes to the design of relay selection strategies.
 In conventional relay  networks, it is well known that the source-relay and
 relay-destination channels are equally important for relay selection, which means that the optimal location of the relay is the middle of the line connecting the
 source and the destination, i.e., ($2.5$ m,0) for
 the scenario considered in Fig. \ref{fig:location}.
Nevertheless,   Fig. \ref{fig:location} shows  that  an EH
relay exhibits   different behavior than a conventional relay, i.e.,
moving the relay from the source towards the middle point ($2.5$ m,0)
has a detrimental   effect on the outage probability.  We note  that this observation is also valid for  SWIPT systems with AF relays.  This phenomenon is due to the fact that
      in EH networks, the quality of the source-relay
     channels is crucial since it  determines not only the transmission reliability from the source to the relays, but also the
     harvested energy at the relays. In \cite{relay_selectionding1}, it was shown  that   the max-min selection criterion, a strategy optimal for conventional
   DF relaying networks, can only achieve a small fraction of the full diversity gain in relaying SWIPT systems.

\section{The Combination of  MIMO  and Cooperative Relaying in SWIPT}
MIMO and cooperative relaying represent two distinct ways of
exploiting   spatial diversity, and both techniques can
significantly enhance the system's energy efficiency, which is of
paramount importance for SWIPT systems. Hence,  the combination of
these  two smart antenna technologies is a natural choice for
SWIPT systems. The benefits of this combination can be illustrated
using  the following example. Consider a lecture hall packed with
students, in which there are many laptops/smart phones equipped
with multiple antennas as well as some low-cost single-antenna
sensors deployed for infrastructure monitoring. This hall can be
viewed as  a heterogeneous network consisting of mobile devices
with different  capabilities. Inactive devices with MIMO
capabilities can be exploited as relays to help the active users
in the network, particularly the low-cost sensors. Since the
relays have multiple antennas, more advanced receiver
architectures, such as antenna switching receivers, can be used.
In addition, the use of these MIMO relays opens the possibility to
serve  multiple source-destination pairs simultaneously.  In this
context, it is important to  note  that   the use of SWIPT will
encourage  the  inactive MIMO users to serve as relays since
helping other users will not reduce the lifetime of the relay
batteries. Therefore, the  MIMO relays can be exploited as an
extra dimension for performance improvement, and  can achieve an
improved trade-off between the   information rate and the harvested
energy \cite{I.Krikidis}.

\begin{figure*}[!h]
\centering
\includegraphics[width=5in]{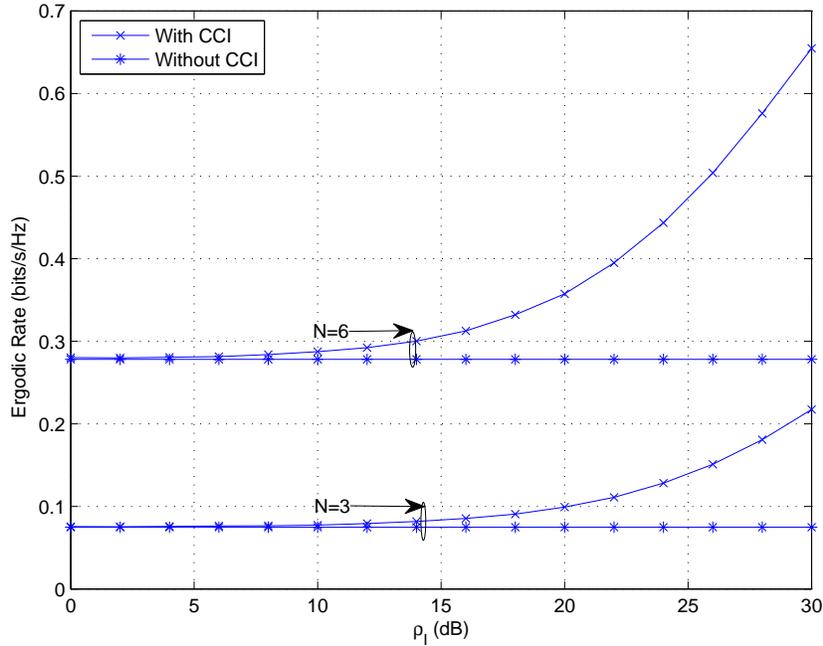}
\caption{\footnotesize Achievable ergodic rate   of a SWIPT relay
system with a single-antenna source, a single-antenna
destination, and a relay with $N$ antennas. The distances from
source to relay, relay and destination, and interferer to relay
are set to 2 m, 3 m, and 5 m, respectively. The path loss exponent
is 3. The total transmit power, noise power, transceiver antenna
gain, and RF-to-electrical energy conversion efficiency are set to
$10$ Watt, $3$ dBm, $0$ dBi, and $80\%$, respectively.   }
\label{fig:CCI}
\end{figure*}

As discussed in Section \ref{section: MIMO SWIPT}, one unique
feature of SWIPT systems is the energy efficient use of CCI, which
is viewed as a detrimental factor that limits   performance in
conventional wireless systems. In particular, CCI  can be
exploited as a potential source of energy in MIMO relay SWIPT
systems. To illustrate this point, let us consider the following example. An AF   relay with $N$ antennas is employed to help a single-antenna source which  communicates  with a single-antenna destination. The relay first harvests energy from the received RF signals with the power splitting  architecture, and then uses this energy  to forward the source signals.
 Two separate cases are
considered, i.e., without CCI and with CCI. To exploit the
benefits of multiple antennas, linear processing of the
information stream is performed to facilitate ID.  Since the optimal linear processing matrix ${\bf W}$ is difficult to characterize analytically, a heuristic rank-1 processing matrix ${\bf W}$ is adopted. As such, in the case without CCI, the processing matrix is designed based on the principle of maximum ratio transmission, i.e., ${\bf W}=a{\bf h}{\bf g}^{\dag}$, where the vectors ${\bf h}$ of size $N\times 1$ and ${\bf g}$ of size $1\times N$ are chosen to match the first and second hop channels, respectively, and $a$ is a scaling factor to ensure the relay transmit power constraint.  On the other hand, in the
presence of CCI, the relay first applies the minimum mean square
error criterion  to suppress the CCI, and then forwards the
transformed signal to the destination using   maximum ratio
transmission. Fig. \ref{fig:CCI} illustrates the achievable
ergodic rate as a function of the average strength of the  CCI
$\rho_I$, with the optimized power splitting ratio. We observe  that increasing the number of relay antennas
significantly improves the achievable rate. For instance,
increasing the number of antennas   from three to six nearly
triples  the rate. Moreover, we see that when the CCI is weak
($\rho_I\leq -10 \mbox{ dB}$), the rate difference is
negligible compared to the case without CCI. However, when the CCI
is strong, a substantial rate improvement is realized. In
fact, the stronger the CCI, the higher the rate gain. For
example, in some applications, the relays will operate at the cell
boundaries and the benefit  of exploiting CCI will be significant
in such situations.

\section{Research Challenges}
In the following, we discuss  some  research challenges for future
MIMO and relay assisted SWIPT.
\begin{enumerate}
\item Energy efficient MIMO SWIPT: Because of severe  path loss
attenuation, the energy efficiency  of MIMO SWIPT systems may not
be satisfactory for long distance power transfer unless advanced
green technologies, such as   EH technologies
relying on natural energy sources,  and MIMO resource allocation
are combined. We now discuss two possible approaches to address
this problem.

\begin{itemize}

\item  EH transmitter: In this case, the
transmitter can harvest energy from natural renewable energy
sources such as solar, wind, and geothermal heat. Then, the
 energy harvested at the transmitter can be transferred   to the
desired receiver over the wireless channel, thereby reducing
substantially the operating costs of the service providers and
improving the energy efficiency of the system, since renewable
energy sources can be exploited virtually for free. However, the
time varying availability of the energy generated from renewable
energy sources may introduce energy outages in SWIPT systems and
efficient new techniques have to be developed to   overcome them.

\item MIMO energy efficiency  optimization: Energy efficient MIMO
resource allocation can be formulated as an   optimization problem
in which the degrees of freedom in the system such as space,  power,
frequency, and time are optimized for maximization of the energy
efficiency.   By taking into account the circuit power consumption
of all nodes, the finite energy storage at the receivers, the
excess spatial degrees of freedom in MIMO systems, and the
utilization of the recycled transmit power and the interference
power, the energy efficiency optimization reveals the operating
regimes for energy efficient SWIPT systems. Yet, the non-convexity
of the energy efficiency objective function \cite{JR:EE_SWIPT}  is
an obstacle in designing algorithms for achieving the optimal
system performance and low-complexity but efficient algorithms are
yet to be developed.

\end{itemize}

\item Energy efficient SWIPT relaying: the concepts of SWIPT and
relaying are synergistic  since the use of SWIPT can stimulate
node cooperation and relaying is helpful to improve the energy
efficiency of SWIPT. In the following, several research challenges
for relay assisted SWIPT   are discussed:
\begin{itemize}

\item
Practical relaying systems suffer from spectral efficiency reduction due
to   half-duplex operation. One possible approach to overcome this limitation   is to use the idea of successive relaying, where two relays listen and transmit in succession. When implemented in a SWIPT system, the inter-relay interference, which is usually  regarded as detrimental, can now be exploited as a source of energy.   Another promising solution is to adopt full-duplex transmission. In the ideal case, full-duplex relaying can double the spectral efficiency, but  the loopback interference corrupts the information signal in practice. Advanced MIMO solutions can be designed to exploit such loopback interference  as an additional source of energy.

\item Relay assisted SWIPT   is not limited to the case of EH relays, and can be extended to   scenarios in which RF EH is performed at the source and/or the destination based on the signals sent by the relay. For example,  in WSNs,     two sensors may communicate with each other with  the  help of a self-powered data fusion center. For this type of SWIPT relaying, the relaying protocol needs to be carefully redesigned, since an extra phase  for transmitting energy to the source and the destination is needed.

\item Most existing works on SWIPT relaying have assumed that all
the energy harvested at the relays can be used as relay
transmission power.  In practice, this assumption is difficult to
realize  due to  non-negligible   circuit power consumption,   power amplifier inefficiency,
 energy storage losses,  and the energy consumed  for   relay network coordination,  which need to be considered
when new SWIPT relaying protocols are designed. In addition,  the
superior performance of MIMO/relay SWIPT is often due to   the key
assumption that perfect CSI knowledge   is available at the
transceivers; however,    a large amount of signalling overhead
will be consumed to realize such CSI assumptions. Therefore, for
fair performance evaluation, future works should take into account
the extra energy cost associated with CSI
acquisition~\cite{Yatawatta}.
\end{itemize}

\item Communication security  management: Energy transfer from the transmitter to the receivers can be facilitated by increasing the transmit power of the information carrying signal. However, a higher transmit power leads to a larger susceptibility for information leakage due to the broadcast nature of wireless channels. Therefore, communication security is a critical issue in systems with SWIPT.

\begin{itemize}
\item Energy signal: Transmitting an energy signal along with the information signal can be exploited for expediting EH at the receivers. In general, the energy signal can utilize arbitrary waveforms such as a deterministic constant tone signal. If the energy signal is a Gaussian pseudo-random sequence, it can  also be used to provide secure communication since it serves as   interference to potential eavesdroppers \cite{JR:Kwan_secure_imperfect}. On the other hand, if the sequence is known to all legitimate receivers, the energy signal can be cancellated at the legitimate receivers before ID. However, to make such cancellation possible, a secure mechanism is needed to share the seed information   for generating the energy signal sequence, to which MIMO precoding/beamforming can be applied.

 \item Jamming is an important means to prevent eavesdroppers from intercepting   confidential  messages; however,
  performing jamming also drains the battery of mobile devices. The use of SWIPT can encourage nodes in a network
  to act as jammers, since they can be wirelessly charged by the RF signals sent by the legitimate users.  However, the efficiency of  this harvest-and-jam strategy  depends on the network topology, where a harvest-and-jam node  needs to be located close to legitimate transmitters  to harvest   a sufficient amount of energy.  Advanced  multiple-antenna
   technologies are needed to overcome this problem.
\end{itemize}

\end{enumerate}

\section{Conclusions}
In this article, the basic concepts of SWIPT and corresponding receiver architectures   have been  discussed  along with some
performance trade-offs in SWIPT systems. In particular,  the application of
smart antenna technologies, such as MIMO and relaying, in SWIPT
systems has  been investigated  for  different  network
topologies. In addition, future research challenges for  the design of energy efficient MIMO and relay assisted SWIPT systems have   been outlined.

 \end{document}